\documentclass[a4paper,11pt]{article}
\pdfoutput=1 

\usepackage{jheppub} 

\usepackage[T1]{fontenc} 

\usepackage{graphicx}
\usepackage{amssymb,amsmath}
\usepackage{color}
\usepackage{lineno}

\title{SOX: Short distance neutrino Oscillations with BoreXino}

\author {
G.~Bellini$^h$,
D.~Bick$^q$,
G.~Bonfini$^e$,
D.~Bravo$^o$,
B.~Caccianiga$^h$,
F.~Calaprice$^k$,
A.~Caminata$^c$,
P.~Cavalcante$^e$,
A.~Chavarria$^k$,
A.~Chepurnov$^p$,
D.~D{\textquoteright}Angelo$^h$,
S.~Davini$^r$,
A.~Derbin$^l$,
A.~Etenko$^g$, 
G.~Fernandes$^c$,
K.~Fomenko$^{b,e}$,
D.~Franco$^{a}$,
C.~Galbiati$^k$,
C.~Ghiano$^a$,
M.~G\"{o}ger-Neff$^m$,
A.~Goretti$^k$,
C.~Hagner$^q$,
E.~Hungerford$^r$,
Aldo Ianni$^e$,
Andrea Ianni$^k$,
V.~Kobychev$^f$,
D.~Korablev$^b$,
G.~Korga$^r$,
D.~Krasnicky$^c$,
D.~Kryn$^a$,
M.~Laubenstein$^e$,
J.M.~Link$^o$,
E.~Litvinovich$^g$,
F.~Lombardi$^e$,
P.~Lombardi$^h$,
L.~Ludhova$^h$,
G.~Lukyanchenko$^g$,
I.~Machulin$^g$,
S.~Manecki$^o$,
W.~Maneschg$^i$,
E.~Meroni$^h$,
M.~Meyer$^q$,
L.~Miramonti$^h$,
M.~Misiaszek$^d$,
P.~Mosteiro$^k$,
V.~Muratova$^l$,
L.~Oberauer$^m$,
M.~Obolensky$^a$,
F.~Ortica$^j$,
K.~Otis$^n$,
M.~Pallavicini$^c$,
E.~Pantic$^s$,
L.~Papp$^o$,
S.~Perasso$^c$,
A.~Pocar$^n$,
G.~Ranucci$^h$,
A.~Razeto$^e$,
A.~Re$^h$,
A.~Romani$^j$,
N.~Rossi$^e$,
R.~Saldanha$^k$,
C.~Salvo$^c$,
S.~Sch\"onert$^m$,
D. Semenov$^l$,
H.~Simgen$^i$,
M.~Skorokhvatov$^g$,
O.~Smirnov$^b$,
A.~Sotnikov$^b$,
S.~Sukhotin$^g$, 
Y.~Suvorov$^{s,g}$,
R.~Tartaglia$^e$,
G.~Testera$^c$,
E. Unzhakov$^l$,
R.B.~Vogelaar$^o$,
H.~Wang$^s$,
M.~Wojcik$^d$,
M.~Wurm$^q$,
O.~Zaimidoroga$^b$,
S.~Zavatarelli$^c$,
and G.~Zuzel$^d$ \\
}

\affiliation[a]{APC, Univ. Paris Diderot, CNRS/IN2P3, CEA/Irfu, Obs. de Paris, Sorbonne Paris Cit\'e, France}
\affiliation[b]{Joint Institute for Nuclear Research, Dubna 141980, Russia}
\affiliation[c]{Dipartimento di Fisica, Universit\`{a} e INFN, Genova 16146, Italy}
\affiliation[d]{M. Smoluchowski Institute of Physics, Jagellonian University, Krakow, 30059, Poland}
\affiliation[e]{INFN Laboratori Nazionali del Gran Sasso, Assergi 67010, Italy}
\affiliation[f]{Kiev Institute for Nuclear Research, Kiev 06380, Ukraine}
\affiliation[g]{NRC Kurchatov Institute, Moscow 123182, Russia}
\affiliation[h]{Dipartimento di Fisica, Universit\`{a} degli Studi e INFN, Milano 20133, Italy}
\affiliation[i]{Max-Plank-Institut f\"{u}r Kernphysik, Heidelberg 69029, Germany}
\affiliation[j]{Dipartimento di Chimica, Universit\`{a} e INFN, Perugia 06123, Italy }
\affiliation[k]{Physics Department, Princeton University, Princeton, NJ 08544, USA}
\affiliation[l]{St. Petersburg Nuclear Physics Institute, Gatchina 188350, Russia}
\affiliation[m]{Physik Department, Technische Universit\"{a}t M\"{u}nchen, Garching 85747, Germany}
\affiliation[n]{Physics Department, University of Massachusetts, Amherst MA 01003, USA}
\affiliation[o]{Physics Department, Virginia Polytechnic Institute and State University, Blacksburg, VA 24061, USA}
\affiliation[p]{Lomonosov Moscow State University Skobeltsyn Institute of Nuclear  Physics, Moscow 119234, Russia}
\affiliation[q]{Institut f\"ur Experimentalphysik, Universit\"at Hamburg, Germany}
\affiliation[r]{Department of Physics, University of Houston, Houston,  TX 77204, USA}
\affiliation[s]{Physics ans Astronomy Department, University of California Los Angeles (UCLA), Los Angeles, CA 90095, USA}

\abstract{

The very low radioactive background of the Borexino detector, its large size, and the well proved capability to detect 
both low energy electron neutrinos and anti-neutrinos make an ideal case for the study of short distance neutrino oscillations with
artificial sources at Gran Sasso. 

This paper describes the possible layouts of $^{51}$Cr ($\nu_e$) and $^{144}$Ce--$^{144}$Pr ($\bar{\nu}_e$) source experiments in Borexino and shows the expected sensitivity to eV mass sterile neutrinos for three possible different phases of the experiment. Expected results on neutrino magnetic moment, electroweak mixing angle, and couplings to axial and vector currents are shown too. }

\begin{document}
\maketitle
\flushbottom

\section{Introduction}
The standard three-flavor neutrino oscillation paradigm has been established by several solar \cite{bib:gallex}\cite{bib:hs}\cite{bib:sno}\cite{bib:bx1}, atmospheric \cite{bib:sk}, accelerator \cite{bib:minos}\cite{bib:t2k} and reactor experiments \cite{bib:kamland}\cite{bib:daya}\cite{bib:reno}. However, long standing anomalies in datasets of different origins lead to the tantalizing hint of the existence of at least one additional sterile state. Well known pieces of this puzzle are the LSND and MiniBoone anomalies \cite{LSND Collab.} \cite{MiniBooNE Collab.},
and, most relevant for this paper, evidence of missing  $\bar{\nu}_e$  or $\nu_e$ in short baseline reactor and radioactive source experiments. 
Although earlier CMBR data seemed to confirm the anomaly and favor a number of relativistic degrees of freedom corresponding to more than three neutrinos \cite{wmap}, recent more accurate data of Planck  \cite{bib:planck} prefers the standard scenario. A fourth neutrino, however, is not excluded.

An important indication comes from a re-evaluation of the $\bar{\nu}_e$ flux from nuclear reactors \cite{Mention} \cite{Huber} and from the re-analysis of a large set of experimental results obtained with detectors located at short distance (10-100 m) from the core. The new calculation, supposedly more accurate, shows that all experiments but one have measured a $\bar{\nu}_e$ flux significantly lower than expected. Although a very recent re-evaluation of the reactor neutrino anomaly in view of the known and large $\theta_{13}$ seemingly reconciles results from reactor neutrinos with the known neutrinos mixing matrix parameters \cite{bib:vogel}, other authors confirm the existence of the anomaly at 2.7$\sigma$ level \cite{bib:kopp}. The second important indication stems from the gallium solar neutrino experiments (Gallex and SAGE \cite{Kaether} \cite{Giunti1}), which have performed measurements with radioactive $\nu_e$ sources made with $^{51}$Cr and $^{37}$Ar, measuring a flux smaller than expected \cite{bib:giunti2012}.

The reactor and gallium anomalies may be explained by oscillations into one or more sterile components. However, regardless of the possible theoretical interpretations, the existence of these anomalies is an experimental problem that must be solved, in one direction or the other, by better and more sensitive experiments.  This paper describes a comprehensive program which will clarify the experimental issue either by firmly confirming or by discarding the existence of the anomaly and, therefore, of new physics, in short distance neutrino oscillations.

A powerful method to probe the anomalies and possibly test conclusively the sterile neutrino(s) hypothesis is to repeat similar source experiments \cite{Grieb} \cite{Giunti2} \cite{Cribier} with a more intense $\nu_e$ (or $\bar{\nu}_e$) source and a larger, better understood, and lower background detector. In this letter we illustrate a three--phases source experiment based on the ultra-low background Borexino detector at LNGS.

\begin{figure}[t]
\begin{center}
\includegraphics[width=0.80\textwidth]{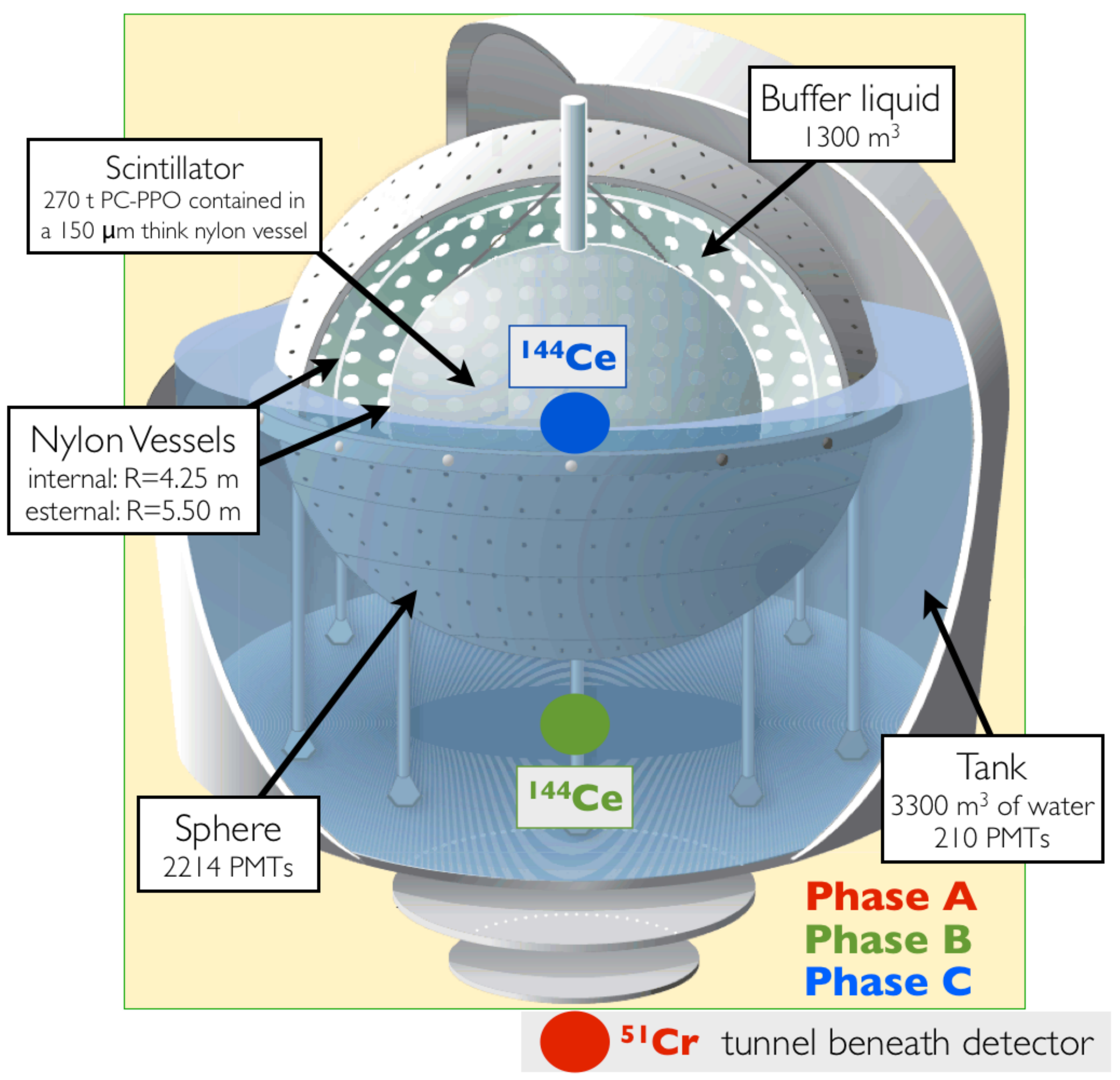}
\caption{\label{fig:detector} Layout of the Borexino detector and the approximate location of the neutrino and anti-neutrino sources in the three phases: Phase A with a $^{51}$Cr neutrino source in a small pit right below the detector center; Phase B with a $^{144}$Ce-$^{144}$Pr anti-neutrino source located right beneath the stainless sphere and within the water tank; finally, Phase C,  with a $^{144}$Ce-$^{144}$Pr anti-neutrino source located inside the scintillator volume. }
\end{center}
\end{figure}

\section{The Borexino detector and the SOX experiment}

Borexino, a large volume ultra-pure liquid scintillator detector, has recently precisely measured several low energy solar neutrino components \cite{Belliniprl107} \cite{Belliniplb707} and has performed the first un-ambiguous detection of geophysical $\bar{\nu}_e$ \cite{bib:geo}. Because of its extremely low background, even the tiny signal of pep solar ${\nu}_e$  could be observed \cite{Belliniprl108}. These features, together with its large radius (up to 11 m of active diameter with $\bar{\nu}_e$), make Borexino a perfect environment for a short distance oscillation experiment.

The experiment (named SOX - Short distance neutrino Oscillations with BoreXino) will be carried out by using in a first instance (Phase A) a $^{51}$Cr $\nu_e$ source of 200-400 PBq activity deployed at 8.25 m from the detector center; in a second phase (Phase B) by deploying a $^{144}$Ce--$^{144}$Pr $\bar{\nu}_e$ source with 2-4 PBq activity at 7.15 m from the detector center, and, finally, in a possible Phase C, a similar $^{144}$Ce--$^{144}$Pr $\bar{\nu}_e$ source located right in the center of the liquid scintillator volume. Fig. \ref{fig:detector} shows a schematic layout of the Borexino detector and the approximate location of the neutrino and anti-neutrino sources in the three phases. 

The Phase C is in principle the most attractive because its sensitivity is definitely higher, as shown later. However, it can  be done only after the conclusion of the solar neutrino program (Borexino Phase 2) and requires a lot of work on the Borexino detector. On the contrary, the Phases A and B, though yielding a slightly lower sensitivity, may be done any time even during the solar neutrino phase of the experiment, which is supposed to continue until the end of 2015, and do not require any change to Borexino hardware. They will not only probe a large fraction of the parameters' space governing the oscillation into the sterile state, but also provide a unique opportunity to test low energy ${\nu}_e$ and $\bar{\nu}_e$ interactions at sub-MeV energy \cite{Berezhiani}. 
Very important, the $^{51}$Cr experiment will benefit from the experience of Gallex and SAGE that in the 90Õs prepared similar sources \cite{Hampel} \cite{SAGE}.

Right beneath the Borexino detector, there is a cubical pit (side 105~cm) accessible through a small squared tunnel (side 95~cm) that was built at the time of construction with the purpose of housing possible neutrino sources. The existence of this tunnel is one of the reasons why the $^{51}$Cr experiment (Phase A) can be done with no changes to the Borexino layout. The center of the pit is at 8.25 m from the detector center, requiring a relatively high source activity for $^{51}$Cr of 200-400 PBq. These values are challenging, but only a factor 2-4 higher than what already done by Gallex and SAGE in the 90s.

In the $^{144}$Ce--$^{144}$Pr experiment the two order of magnitude lower attainable activity suggests to deploy the source both externally at 7.15 m from the center (within the water tank, Phase B) or, even better, within the detector itself (Phase C). The activity of the source in these cases should be 2.3 PBq for the external source and about 1.5 PBq for the internal one. In both cases, the sensitivity can be enhanced by inserting PPO \cite{bib:liquid} in the buffer liquid, in order to increase the scintillator radius for the 
detection of $\bar{\nu}_e$ from the current 4.25 m up to 5.50 m \cite{Alimonti}. 

The challenge for the Phase C is constituted by the large background induced by the source in direct contact with the scintillator, that can be in principle tackled thanks to the correlated nature of the $\bar{\nu}_e$ signal detection. In Phase B this background, though still present, is mitigated by the shielding of the buffer liquid.

\section{Neutrino and anti-neutrino sources}

Neutrinos are detected in Borexino by means of the scintillation light emitted by scattered electrons and the electron energy is reconstructed from the total amount of light. The light pulse is a few 100 ns long and pulse shape analysis allows disentangling $\beta$-like events from $\alpha$-like events very efficiently \cite{bib:pulse-shape}. The position of each event is reconstructed by time--of--flight with a resolution of about 15 cm at 0.7 MeV, so that the distance from the source of each event can be known with that precision.

Anti-neutrinos are neatly and efficiently detected by means of inverse beta decay (IBD) on protons. The low radioactivity and the clean tag offered by the space-time coincidence between the prompt $e^+$ and the subsequent neutron capture 
($\tau$=254 $\mu$s \cite{bib:geo}) make accidental background essentially zero (less than 1 event per year in the total volume). The detection threshold for IBD is 1.8 MeV, matching adequately the energy of the $\bar{\nu}_e$ emitted by the $^{144}$Ce--$^{144}$Pr source (end point about 3 MeV).

The $^{51}$Cr source will be produced by irradiating a large sample of highly enriched $^{50}$Cr in a nuclear reactor which may accommodate such a large volume of Cr and yield a high thermal neutron flux ($\approx$10$^{15}$ cm$^{-2}$ s$^{-1}$). The amount of Cr may vary from 10 kg up to 35 kg, depending on the level of enrichment. 
The $^{144}$Ce--$^{144}$Pr source is done by chemical extraction of Ce from exhausted nuclear fuel \cite{Cribier}. $^{51}$Cr decays via electron capture into $^{51}$V, emitting two neutrino lines of 750 keV (90\%) and 430 keV (10\%), while $^{144}$Pr
decays $\beta$ into $^{144}$Nd with an end--point of 3 MeV ($^{144}$Ce decays $\beta$ too, but is below threshold).    

\begin{figure}[t]
\begin{center}
\includegraphics[angle=90, width=0.95\textwidth]{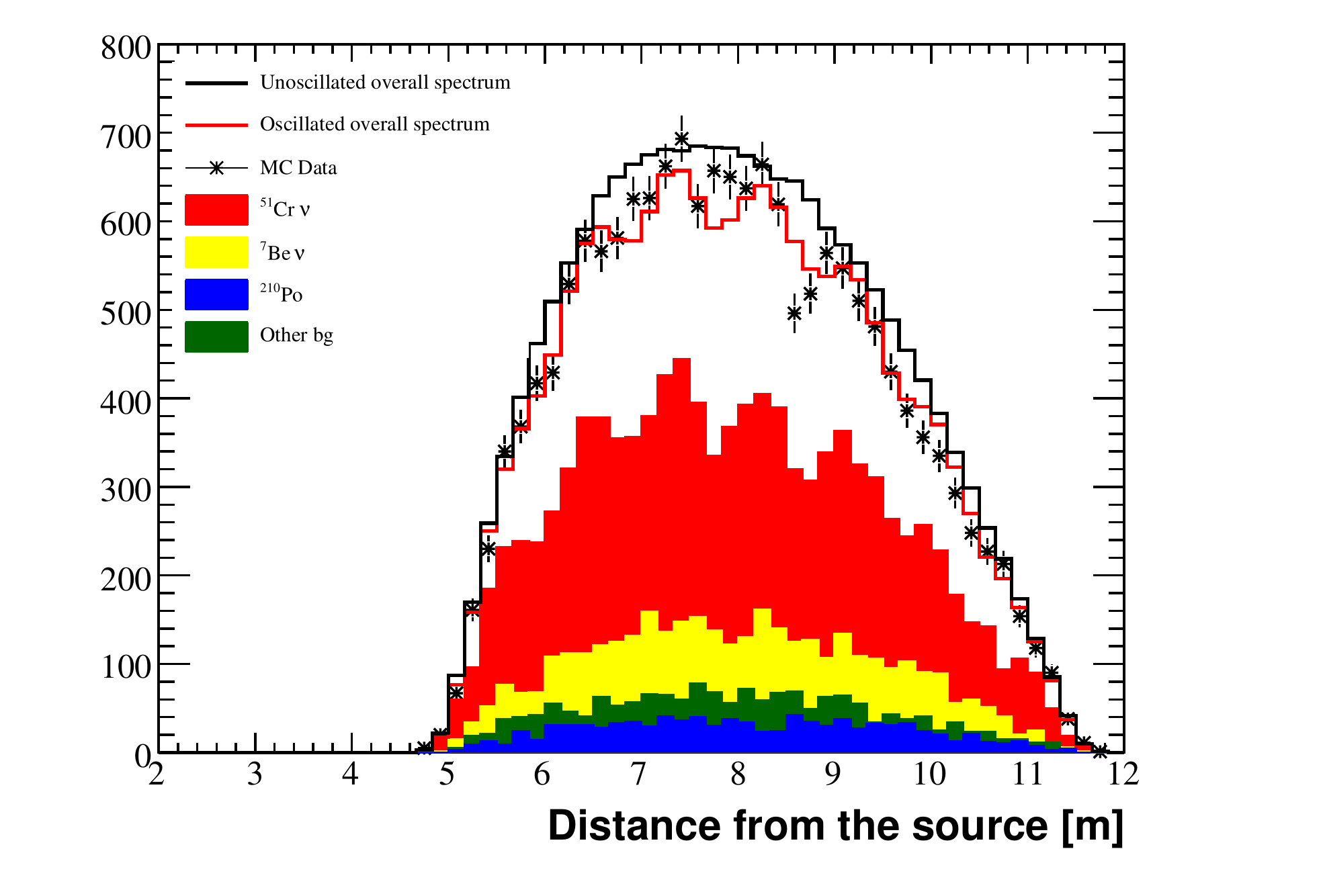}
\caption{\label{fig:cr51-waves} Example of a possible outcome of the $^{51}$Cr experiment (Phase A) with $\sin^2(2\theta_{14})$=0.3 and $\Delta m_{41}^2$=2 eV$^2$. Data points are obtained with a full Geant-4 simulation that was validated at 1\% level with calibration sources, including known backgrounds. The signal (red band) is dominating at all distances from the source.  The oscillatory behavior allows to reconstruct $\theta_{14}$ and $\Delta m_{41}^2$. }
\end{center}
\end{figure}

Borexino can study short distance neutrino oscillations in two ways. The first way is the standard disappearance technique used by many experiments at reactors, accelerators, and with solar neutrinos: if oscillations occur, the total count rate is lower
than that expected without oscillations. The second way relies on the idea to perform an ``oscillometry'' measurement within the detector volume \cite{Grieb}: due to the fact that the values of $\Delta m_{41}^2$ inferred from the existing neutrino anomalies is of the order of 1 $eV^2$ and that the energy of radioactive induced neutrinos is of the order of 1 MeV, the typical oscillations length amounts to a few m and the resulting oscillations waves can be directly ``seen'' with a large detector like Borexino. This is easily understood from the well-known two-flavor oscillation formula:

\begin{center}
$P_{ee}=1-  \sin^{2}2 \theta_{14}  \sin^{2} \frac{1.27  \Delta  m_{41}^{2} ( eV^{2}) L(m) }{E (MeV)} $\\
\end{center}

where $\theta_{14}$ is the mixing angle of the $\nu_e$ (or $\bar{\nu}_e$) into sterile component, $\Delta m_{41}^2$ is the corresponding squared mass difference, L is the distance of the source to the detection point, and E is the neutrino energy. The imprinting of the survival probability $P_{ee}$ on the spatial distribution of the detected events is shown in Fig.~\ref{fig:cr51-waves}, for the $^{51}$Cr source (Phase A), indicating that for appropriate values of $\theta_{14}$ and $\Delta m_{41}^2$ the oscillometry behavior is clearly detectable, if exists, as waves superimposed on the event profile in space. It is evident from the figure that the experiment would also be sensitive to any deformation of the shape. Oscillation parameters can be directly extracted from the wavelength and amplitude of the wave. 

This result may be obtained only if the size of the source is not too big. The $^{51}$Cr source will be made by about 10-35 kg of highly enriched Cr metal chips which have a total volume of about 4-10 l. The real weight and volume will depend on the final level of isotopic enrichment of the material which may vary from 38\% up to maximum 95\%. The source linear size will be therefore about 15-23 cm, comparable to the position reconstruction resolution of the events. The $^{144}$Ce--$^{144}$Pr
source is made of a few hg of Ce and its size is negligible. In all simulations shown below the source size effect is included.

\begin{figure}[t]
\begin{center}
\includegraphics[width=0.95\textwidth]{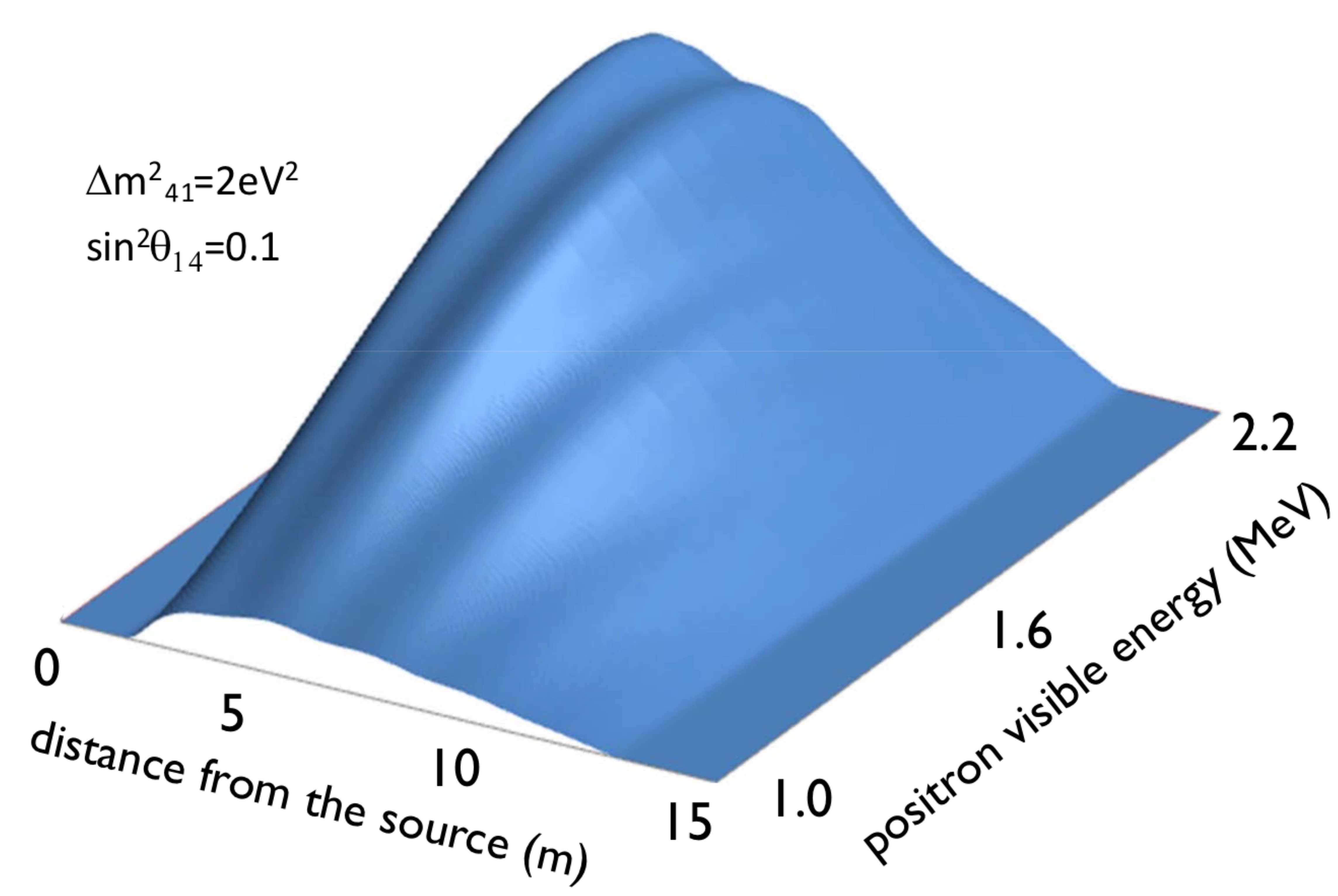}
\caption{\label{fig:ce144-waves} The oscillometry pattern as a function of the reconstructed (positron) visible energy for the Phase B experiment. A similar distance--energy correlation is expected in Phase C. }
\end{center}
\end{figure}

Fig.~\ref{fig:ce144-waves} shows the same pattern for Phase B in a 3D plot, where the correlation between the waves and the reconstructed  $\bar{\nu}_e$ energy is evident. This pattern is very powerful and allows to reconstruct $\theta_{14}$ and $\Delta m_{41}^2$ even if the $\bar{\nu}_e$ spectrum of the $^{144}$Ce--$^{144}$Pr is not mono-chromatic. 

In Phase A the total counts method sensitivity is enhanced by exploiting the fact that the life-time of the $^{51}$Cr is relatively short. The known time-dependence of the signal, and the concurrent assumption that the background remains constant along the measurement (a fact that we know from Borexino data) significantly improves the sensitivity.  In Phases B and C this time-dependent method is not effective because the source life-time is longer (411 days), but this is more than compensated by the very low background and by the larger cross-section.

The total counts and waves methods combined together yield a very good sensitivity for both experiments. Besides, the wave method is independent on the intensity of the source, on detector efficiency, and is potentially a nice probe for un-expected new physics in the short distance behavior of neutrinos or anti-neutrinos. 

\section{SOX sensitivity to sterile neutrinos}

In the following, we report the sensitivity of the three phases. The sensitivity plots are compared with the contours of the reactor anomaly, and also with the mixing angle upper bound obtained from the solar neutrino experiments and from the relatively large value of $\theta_{13}$. 

For the $^{51}$Cr experiment we assume to achieve 1\% error in the measurement of the source activity, and 1\% error in the knowledge of the fiducial volume with which we select the candidate events. The first number is challenging, but feasible (in 1995 Gallex experiment obtained about 2\% precision). Preliminary analysis has shown that a precision of 1\% in the activity might be obtained with a carefully designed and precisely calibrated isothermal calorimeter in which the activity is measured through a very precise knowledge of the heat released by the source. The calorimeter will be designed to allow the calorimetric measurement both during and after the data taking. 

The second one is even conservative: with a careful calibration by means of standard sources (already foreseen for solar physics), the achievement of better than 1\% knowledge of fiducial volume is realistic \cite{Belliniprl107}. 

For the $^{144}$Ce--$^{144}$Pr experiments we assume instead a 1.5\% source intensity precision; furthermore, due to the correlated nature of the signal, we do not consider applying a fiducial volume cut (the whole scintillator coincides with the active volume) and therefore we omit the corresponding error. However, we include a 2\% experimental systematic error, uncorrelated between energy and space bins, to account for residual systematic effects.

The sensitivity of the 370 PBq $^{51}$Cr source test is evaluated  assuming to deploy the source in the tunnel under the detector (8.25 m from the center) and after 6 days from its activation. The advantage of this proposal is that in Borexino the background is already very accurately measured and known, since the source is not in contact with or close to the active mass, and does not induce any further contamination. The only expected background components are solar neutrinos (mostly $^{7}$Be, which Borexino has accurately measured) and small amounts of radioactive contaminants  intrinsic to the scintillator. The latter, in particular, are mostly due to the sizable $^{210}$Po ($\tau$ = 199.6 days) content of the scintillator, for which we predict at the time of the test (around 2015) a rate of about 11.8 cpd in 133 tons of scintillator (selected as fiducial volume for the measurement), and in the energy region of interest, [0.25--0.7] MeV. The constant background is due to long living ($>$10 y) isotopes intrinsic to the scintillator, like $^{210}Pb$ (through the daughter $^{210}$Bi) and $^{85}$Kr, to solar neutrinos, and to gammas from the detector materials. The overall constant background rate is estimated in 54 cpd/133 ton of scintillator, taking into account the recent achievement in the purification of the scintillator 
(this background was higher in \cite{Belliniplb707}\cite{bib:geo}\cite{Belliniprl108}). We assume, conservatively, a detector duty cycle equals to 90\%, and a data taking with the source for 100 days. The signal is extracted by looking for the $^{51}$Cr decay mean life (39.966 days) and the oscillation induced distortion in the spatial distribution.

\begin{figure}[t]
\begin{center}
\includegraphics[width=0.95\textwidth]{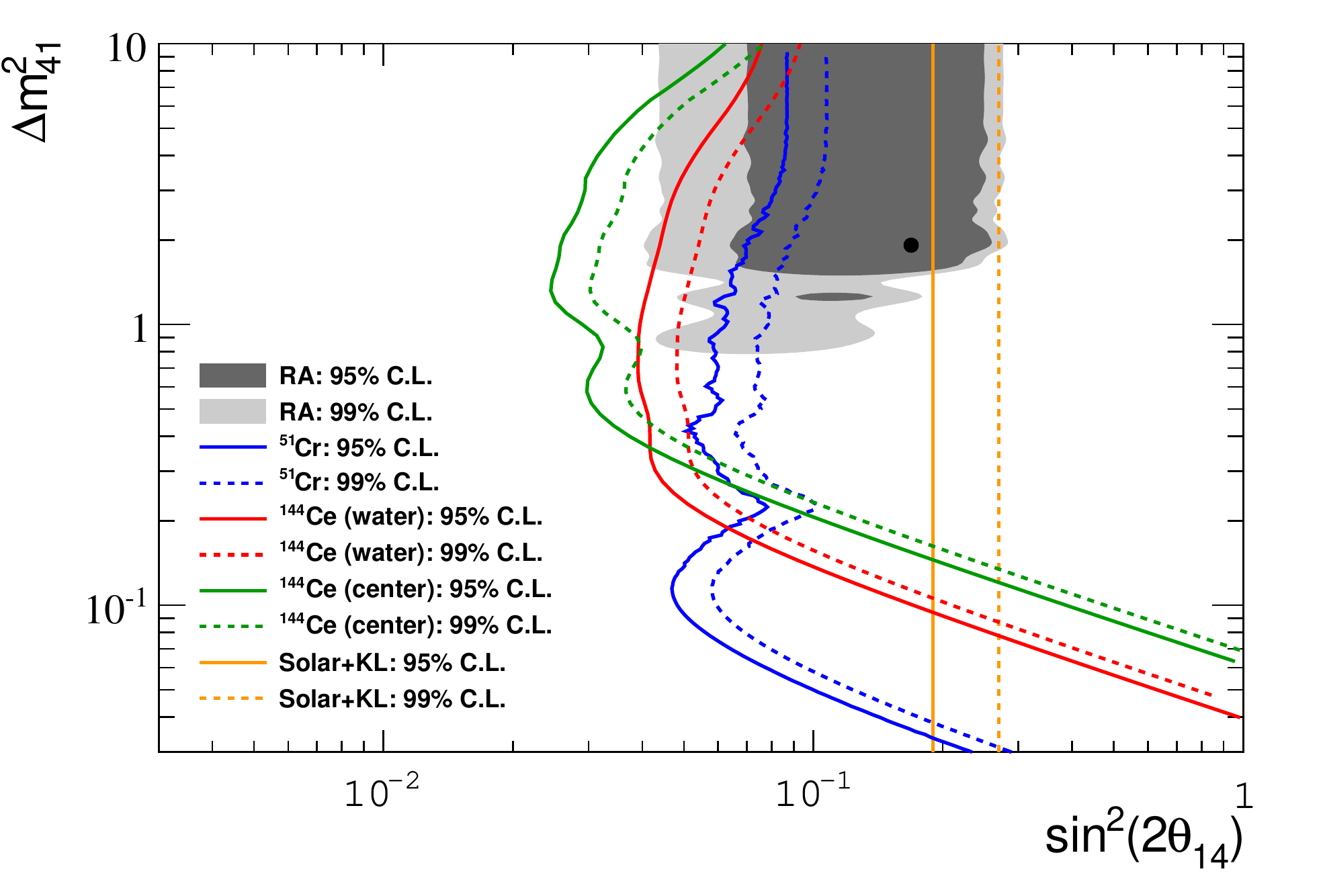}
\end{center}
\caption{\label{fig:sensitivity} Sensitivity of the Phase A ($^{51}$Cr external, blue), of Phase B ($^{144}$Ce--$^{144}$Pr external, red) and Phase C ($^{144}$Ce--$^{144}$Pr center, green). The grey area is the one indicated by the reactor anomaly, if interpreted as oscillations to sterile neutrinos. Both 95\% and 99\% C.L. are shown for all cases. The yellow line indicates the region already excluded in \cite{bib:palazzo2}. }
\end{figure}

The sensitivity is evaluated with a toy Monte Carlo approach, which consists in generating 2000 data samples for each pair of  
$\Delta m_{41}^2$ and $sin^2(2 \theta_{14})$ parameters,  according to the expected statistics.  In the simulation, we assume a period of 15 weeks of stable data taking before the source insertion in order to accurately constrain the background. The analysis model includes all the known background components, which have been accurately studied and measured from 2007 until now for solar physics,  and the non-oscillated signal. We built the confidence intervals from the mean $\chi^2$ for each couple of parameters with respect to the non-oscillations. The result is shown in Fig.~\ref{fig:sensitivity}. It is evident from the figure that the reactor anomaly region is mostly covered. 

Our baseline plan is to reach this sensitivity with a single irradiation of 370 PBq (10 MCi).  A similar result can be obtained with two irradiations of about 200 PBq. A single irradiation is preferable and yields a slightly better signal to noise ratio. The two-irradiations option, however, is acceptable and yields very similar results. We remind that, because of the un-avoidable $\gamma$ background from the source, we cannot put the $^{51}$Cr source inside.

The physics reach  for the $^{144}$Ce--$^{144}$Pr external (Phase B) and internal (Phase C) experiments, assuming 2.3 PBq (75 kCi) source strength and one and a half year of data taking) is shown in Fig.~\ref{fig:sensitivity}.  
The $\chi^2$ based sensitivity plots are computed assuming an active radius of 5.5 m, compared to the current active radius of 4.25 m for the solar phase. Such an increase will be made possible by the addition of the scintillating fluor (PPO) in the inner buffer region (presently inert) of the detector.
We have also conservatively considered to blind in the analysis a sphere centered in the origin and of 1.5 m radius to reject the gamma and bremsstrahlung backgrounds from the source assembly itself. Under all these realistic assumptions, it can be noted from Fig.~\ref{fig:sensitivity} that the intrinsic $^{144}$Ce--$^{144}$Pr sensitivity is very good: for example the 95\% C.L. exclusion plot predicted for the external test covers adequately the corresponding reactor anomaly zone, thus ensuring a very conclusive experimental result even without deploying the source in the central core of the detector. The background included in the calculation is negligible, being represented by about 5 $\bar{\nu}_e$ events per year from the Earth (geo-neutrinos) and from distant reactors, with negligible contribution from the accidentals. It is worth to stress that the three ingredients at the origin of this good performance are the very low background due to the $\bar{\nu}_e$ coincidence tag, the larger cross-section due to the higher source energy, and the deployment of the source closer or directly within the active volume detector, yielding a larger geometrical acceptance. 

\begin{figure}[t]
\begin{center}
\includegraphics[width=0.95\textwidth]{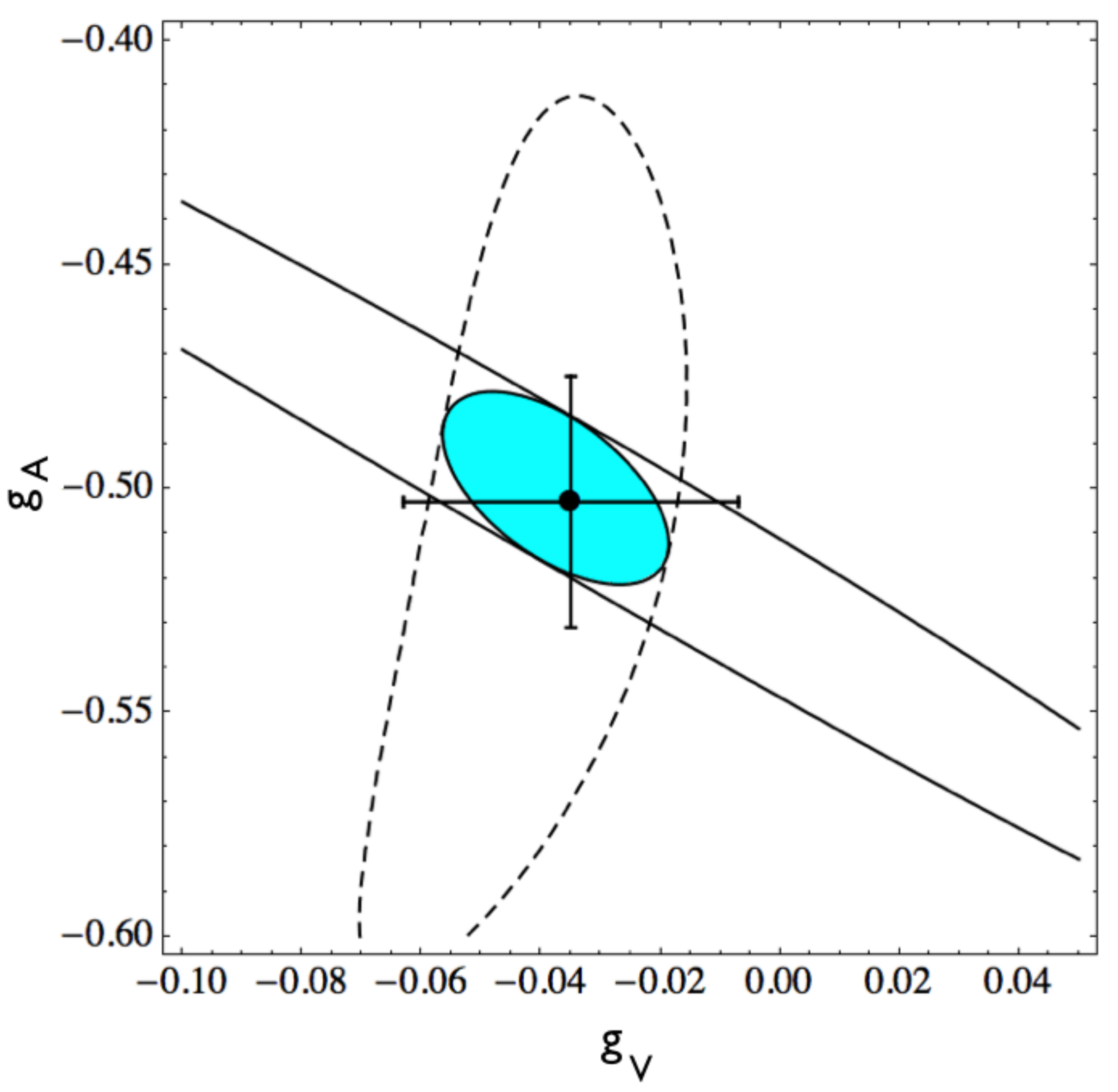}
\caption{\label{fig:gvga}Sensitivity to the measurement of g$_V$-g$_A$ by using the data of Phase A and Phase C. }
\end{center}
\end{figure}

\section{Other physics goals}

SOX will yield additional physics. The electroweak angle $\theta_W$ can be directly measured at MeV scale from the ${\nu}_e$-e$^-$ cross--section with an expected precision of 2.6\%. This value is better that any other obtained at this energy scale. Furthermore, Phase A will provide significant information about neutrino magnetic moment \cite{Ianni}\cite{Fiorentini} and improve the best result obtained so far~\cite{Beda}.

By combining the ${\nu}_e$-e$^-$ of Phase A and $\bar{\nu}_e$-p data of Phase C the vector (g$_V$) and axial (g$_A$) current coefficients of the low energy Fermi current-current interaction can be measured. In the standard model g$_V$=$-\frac{1}{2}+2\sin^2\theta_W$ and g$_A$=$-\frac{1}{2}$. The best measurement at relatively low energy (10 GeV) was  obtained by the CHARM II experiment \cite{bib:charm2}. As shown in Fig. \ref{fig:gvga} Borexino can obtain a similar (actually a little better) precision at much lower energy, where the existence of additional non-standard interactions might more easily probed (the Fermi cross section grows with energy, so the standard model interaction in Borexino will be three orders of magnitude smaller than in CHARM 2).

\section{Conclusions}
In summary, Borexino is an ideal detector to test the existence of sterile neutrinos through the identification of the disappearance/waves effects induced by the mixing with the active states.  The staged approach proposed in this paper, first envisioning an external $^{51}$Cr source and later two $^{144}$Ce--$^{144}$Pr experiments, is a comprehensive sterile neutrino search which will either confirm the effect or reject it in a clear and unambiguous way. Particularly, in case one sterile neutrino with parameters corresponding to the central value of the reactor anomaly, SOX will surely discover the effect, prove the existence of oscillations and measure the parameters through the ``oscillometry'' analysis.

\acknowledgments
Borexino was made possible by funding from INFN (Italy), NSF (USA), BMBF, DFG, and MPG (Germany), NRC Kurchatov Institute (Russia), MNiSW (Poland, Polish National Science Center (grant DEC-2012/06/M/ST2/00426)), and Russian Foundation for Basic Research (Grant 13-02-92440 ASPERA). We acknowledge the generous support of the Gran Sasso National Laboratories (LNGS). SOX is funded by the European Research Council.

\end{document}